\documentclass[a4paper]{jpconf}
\usepackage{graphicx}
\begin{document}
\title{Equation of State within Gluon Dominated QGP Model in Relativistic Hydrodynamics Approach}

\author{T. P. Djun}

\address{Graduate Study in Material Science, 
  University of Indonesia, 
  Kampus UI Salemba, \\
 Jakarta 10430, Indonesia \\
Group for Theoretical and Computational Physics,
  Research Center for Physics, \\ 
  Indonesian Institute of Sciences, Kompleks Puspiptek Serpong, \\
  Tangerang 15310, Indonesia}

\ead{tp.djun@sci.ui.ac.id}

\author{M. K. N. Patmawijaya}

\address{Departemen Fisika, FMIPA, Universitas 
  Indonesia, \\
  Depok 16424, Indonesia}
  
\author{R. Utama}

\address{Departemen Fisika, FMIPA, Universitas 
  Indonesia, \\
  Depok 16424, Indonesia}

\author{L. T. Handoko}

\address{Group for Theoretical and Computational Physics,
  Research Center for Physics, \\
  Indonesian Institute of Sciences, Kompleks Puspiptek Serpong, \\
  Tangerang 15310, Indonesia}

\begin{abstract}
The dynamics of Quark-gluon plasma (QGP) as a lump of deconfined free quarks and gluons is elaborated. Based on the first principal we construct the Lagrangian that
represents the dynamics of QGP. To induce a hydrodynamics
approach, we substitute the gluon fields with flow fields. As a result, 
the derived equation of Motion (E.O.M) for gluon dominated QGP shows 
the form that similar to Euler equation, and the energy momentum tensor 
also represents explicitly the system of ideal fluid. Combining the E.O.M and 
energy momentum tensor, the pressure and energy density distribution as the equation of states are analytically derived. 
\end{abstract}

\section{Introduction}

Recent experiments on heavy-ion collisions show a strong indication 
that hot dense deconfined phase of free quark and gluon, the so called 
quark-gluon plasma (QGP), is conjectured to exist. 
The study of QGP itself has been carried out 
through a number of different approaches in the previous works. 
Some results of these studies were obtained in the framework of quantum 
chromodynamics (QCD) theory by utilizing the lattice gauge  calculation 
\cite{gottlieb,petreczky}. Other calculations of QGP were
based on the relativistic hydrodynamics approach \cite{bouras,romatschke}. 
In the latter, the QGP could  be either quark \cite{romatschke} or gluon 
\cite{bouras} dominated matter. For the quark dominated approach, 
a very small ratio of shear viscosity over entropy is required to 
get a good fit of the spectra of transverse momentum, energy density 
distribution and other physical observables that are obtained from 
experiments \cite{teaney,huovinen,kolb,kolb1,hirano,baier}. 
On the other hand, the gluon dominated plasma motivated by 
the discoveries of jet quenching in the heavy-ion collision at RHIC 
indicates the shock waves in the form of Mach cone \cite{adams,adare}. 
The present paper adopts the so-called 
fluid QCD model \cite{sulaiman,djun} to produce the equation of 
motion and energy momentum tensor for quark and gluon in a lump of QGP,
and subsequently to investigate the distribution of the pressure and 
energy density. 

This paper is organized as follows. In Section \ref{sec:model} the 
fluid QCD model is briefly revisited, and the energy momentum tensor 
for ideal fluid is derived. Then, it is followed by the derivation
for the equation of state and the explicit expression of gluon field in Section \ref{sec:field_equation}.
Finally, the summary and discussion will be given in Section 
\ref{sec:summary_and_discussion}.
Throughout this work, we use the natural units, i.e, $\hbar \equiv c \equiv 1$.

\section{Model}
\label{sec:model}

The Lagrangian of QGP that describes  the unification of fermions and 
bosons from different gauge groups with preserving ${\rm SU}(3)_{F} \otimes {\rm U}(1)_{G}$ 
gauge symmetry can be written as  \cite{tpdjun,sulaiman}
\begin{equation}
{\cal L} = i \bar{Q} \gamma^\mu \partial_\mu  Q - m_Q \bar{Q} Q - 
{\textstyle \frac{1}{4}} S^a_{\mu\nu} S^{a\mu\nu}  - {\textstyle \frac{1}{4}} F_{\mu\nu} F^{\mu\nu} + g_{F} 
J^a_{F \mu} U^{a\mu} + g_G J_{G \mu} A^{\mu} ,
\label{eq:Lagrangian-ideal}
\end{equation}
where $Q$ and $\bar{Q}$ represent the quark and anti-quark triplet, 
$\gamma^\mu$ is Dirac gamma matrices, and $m_Q$ is the mass of quark. 
The factor 
${\textstyle \frac{1}{4}} S^a_{\mu\nu} S^{a\mu\nu}$ is the gauge invariant 
kinetic term of gluon field. The gluon field strength tensor itself 
is expressed as
$S^a_{\mu\nu} \equiv \partial_\mu U^a_\nu - 
\partial_\nu U^a_\mu + g_F f^{abc} U^b_\mu U^c_\nu$. 
In the latter, $U^a_{\mu}$ indicates the gluon field, 
$g_F$ is the strong coupling constant, and $f^{abc}$
is the structure constant of SU(3) gauge group. 
The kinetic term for gauge boson $A_\mu$ is built inside its field 
strength tensor, i.e.,
$F_{\mu\nu} \equiv  \partial_\mu A_\nu - \partial_\nu A_\mu $. 
The last two terms of Eq.~(\ref{eq:Lagrangian-ideal}), 
$J^{a}_{F \mu} = \bar{Q} T^a_F 
\gamma_\mu Q$ and $J_{G \mu} = \bar{Q} \gamma_\mu Q$, represent the quark 
currents from the SU(3) and U(1) gauge groups, respectively, 
whereas $g_G$ denotes 
the coupling constant from U(1) gauge group.

The QCD Lagrangian in Eq.~(\ref{eq:Lagrangian-ideal}) is constructed 
with a purpose to reproduce the energy momentum tensor that has 
the same form as the energy momentum tensor of an ideal fluid \cite{tpdjun}.  
In the succeeding step, the gluon field $U^a_{\mu}$, 
that is designed  to act as the flow field in the system is proposed to be 
formulated in a configuration that inherently contains 
the relativistic flow. 
It is formulated as  $U^a_\mu = (U^a_0 , \textbf{U}^a) \equiv u_\mu \phi^a$ 
\cite{sulaiman,djun}. 
Here, $\phi^a$ is the dimension one scalar field, 
and $u_\mu \equiv \gamma  (1, \textbf{v} )$ is the
relativistic velocity, with $\gamma  = (1-\vert\textbf{v} \vert^2)^{-1/2}$.

This formulation provides a clue that a single gluonic field $U^a_\mu$
may behaves as a fluid at certain scale, beside its conventional point particle
properties with a polarization vector $\epsilon_\mu$ in the form of  $U^a_\mu =
\epsilon_\mu \, \phi^a$. One can then consider that there is a kind of ``phase
transition'',
\begin{eqnarray}
  \underbrace{\mathrm{hadronic \; state}}_{\displaystyle \epsilon_\mu} \longleftrightarrow \underbrace{\mathrm{QGP \; state}}_{\displaystyle u_\mu} . \nonumber
\end{eqnarray}

As the gluon field behaves as a point particle, it is in a stable hadronic state and is characterized by its polarization vector. On the other hand in the pre-hadronic state (before hadronization) like hot QGP, the gluon field  behaves as a highly energized flow particle and the properties are dominated by its relativistic velocity. 

The field $U^a_\mu$ is actually analogous to the gauge boson from $U(1)$ gauge group in particle physics, where the polarization vector $\epsilon_\mu$ from the free photon solution $\sim \epsilon_{\mu}  \exp(-i p_{\mu} x^{\mu})$ is replaced by  the 4-velocity $u_{\mu}$. Recall that the wavefunction $U^a_{\mu}$ for a free particle satisfies $[g^{\mu\nu} (\partial^2 + m^2 ) - \partial^{\nu} \partial^{\mu}] U^a_{\mu} = 0$ with solution $U^a_{\mu} \sim \epsilon_{\mu} \;\phi^a \exp(-i p_{\mu} x^{\mu})$. For a massive vector particle, $m  \neq 0$, we have no choice but to take $\partial^{\mu} U^a_{\mu} = 0$. It is not a gauge condition like the case of massless particle. This then demands $p^{\mu} \epsilon_{\mu} = 0$. The number of independent polarization vectors is reduced from four to three in a covariant fashion. However, one can still perform another gauge transformation to the massless $U_\mu^a$ which  makes finally only two degrees of freedom remain. Therefore, one should keep in mind that in the present model the spatial velocity has only two degrees of freedom, that means one component must be described by another two vector components.\\
Further, when Euler-Lagrange equation is applied to 
Eq.~(\ref{eq:Lagrangian-ideal}) one obtain
\begin{equation}
\frac{\partial}{\partial t} (\gamma \mathbf{v} \phi^a) + \nabla 
(\gamma \phi^a) = -g_F \oint d\mathbf{x} (J^a_{F 0} + F^a_0) ~,
\label{eq:eq-of-motion}
\end{equation}
where
$J^a_{F0}$ denotes the covariant current originating from the quarks 
that are surrounded by and interact with the 
gluon "fluid", while the term $F^a_\mu$ is 
induced by the fluid self-interaction and the interacting 
gauge fields $A^a_\mu$.
Equation~(\ref{eq:eq-of-motion}) is considered as the general relativistic 
fluid equation for single gluon field $U^a_{\mu}$, since in the 
non-relativistic limit, i.e. $\phi^a \sim 1$ and $\gamma \sim 1 
+ \frac{1}{2} \vert \mathbf{v} \vert^2 $,
Eq.~(\ref{eq:eq-of-motion}) transforms to the classical equation of 
motion of fluid dynamics \cite{sulaiman}

\begin{equation}
\frac{\partial \mathbf{v} }{\partial t} + (\mathbf{v} \cdot \nabla) 
\mathbf{v} = -g_F \oint d\mathbf{x} (J^a_{F 0} + F^a_0)\mid_{\rm non-rel} ~.
\label{eq:eq-of-motion-nonrel}
\end{equation} 
This shows that from a certain point of view the Lagrangian describes 
a general relativistic fluid system interacting with another gauge 
field and the matter inside.  
The flow characteristic that appears in the equation of motion comes
from the contribution of gluon field $U^a_{\mu}$. This fact indicates that the 
system we are working with is a gluon dominated QGP. 
In such system the terms that do not contain gluon field may be omitted.
As a consequence we have
\begin{eqnarray}
{\cal L} &=& - {\textstyle \frac{1}{4}} S^a_{\mu\nu} S^{a\mu\nu}  + g_F J^a_{F \mu} U^{a \mu}  ~.
\label{eq:qgp-Lagrangian}
\end{eqnarray}
The Lagrangian given by Eq.~(\ref{eq:qgp-Lagrangian}) 
describes the kinetics of gluons, 
the self-interaction of gluons, the self-interaction between small number 
of quark and anti-quark, and the interaction between quark with 
gluon ''fluid''. 
Electromagnetic interaction also exist in the system, but the scale is
suppressed due to its tiny value compared with the strong interaction.
From the Lagrangian given in Eq.~(\ref{eq:qgp-Lagrangian}) we 
can derive the energy momentum tensor \cite{hobson}
\begin{eqnarray}
\cal{T}_{\mu\nu} &=& 
\frac{2}{\sqrt{-g}} \frac{\delta ( {\cal L} \sqrt{- g})}{\delta 
g^{\mu\nu}}\nonumber\\
&=& S^a_{\mu\rho} S^{a \rho}_{\;\;\;\; \nu} - g_{\mu\nu} {\cal L} + 2 g_F J^a_{F \mu} U^a_\mu \nonumber\\
&=& [  2 g_F g_{\mu\nu} J^a_{F \mu} U^{a \mu} +  g^2_F f^{abc} f^{ade} U^b_\mu U^c_\rho U^{d \rho} U^e_\nu ] \nonumber \\
&& - [ g_F g_{\mu\nu}  J^a_{F \mu} U^{a\mu} -  \frac{1}{4} g_{\mu\nu} g^2_F f^{abc} f^{ade} U^b_\mu U^c_\nu U^{d \mu} U^{e \nu} ] ~.
\label{eq:energy-momentum-tensor}
\end{eqnarray}
The factor $J^a_{F \mu} U^{a\mu}$ in Eq.~(\ref{eq:energy-momentum-tensor})
might be expressed in a more elementary form. To this end, we
can start with the assumption that the solution of the Dirac 
equation for single color quark 
is in the form of $Q (p, x) = q (p) \exp(-ip \cdot x)$. 
By inserting this solution to the Dirac equation, i.e.,
$(i \gamma^{\mu} \partial_{\mu} - m) q (p) \exp(-ip \cdot x) = 0$,
and multiplying with the anti-quark solution,
we obtain $\bar{q} \gamma_\mu q = 4 p_\mu$.
Since $J^a_{F \mu} = \bar{Q} \gamma_{\mu} Q T^a = \bar{q} \gamma_{\mu} q T^a = 4 p_{\mu} T^a$, 
and $U^{a \mu} = u^\mu \phi^a$, we finally obtain 
$J^a_{F \mu} U^{a\mu} = 4 m_Q T^a \phi^a$.
Note that we have utilized $u_{\mu} u^{\mu} = 1$  along with the assumption 
that all quarks and anti-quarks have the same momenta $P^{\mu}$ 
and that the velocities of all gluons and quarks are homogeneous.
Thus, the energy momentum tensor in the function of field $\phi^a$ reads
\begin{eqnarray}
\cal{T}_{\mu\nu } &=& [ 8 g_F m_Q T^a \phi^a + g^2_F f^{abc} f^{ade} \phi^b \phi^c \phi^d \phi^e  ] u_\mu u_\nu \nonumber \\
&&- [ 4 g_F m_Q T^a \phi^a  -  \frac{1}{4} g^2_F f^{abc} f^{ade} \phi^b \phi^c \phi^d \phi^e ] g_{\mu\nu} ~.
\label{eq:energy-momentum-tensor_ideal}
\end{eqnarray}
With this form, ${\cal T}_{\mu\nu}$ is obviously describing a system of  perfect fluid. 
Further, we assume that the gluon color states are 
homogeneous, i.e., $\phi^a = \phi$ for all $a = 1,\cdots,8$.
As the follow-on to this assumption, 
the generator $T^a$ can be compactly written as $T = \sum_{a} T^a$
and the 2nd and 4th 
terms of $\mathcal{T}_{\mu\nu}$ can be omitted due to the completely 
anti-commutative property of the structure constant $f^{abc}$.
The energy momentum tensor then reduces to
\begin{eqnarray}
\cal{T}_{\mu\nu}  &=& (8 T g_F m_Q \phi) u_\mu u_\nu -  (4 T g_F m_Q \phi) g_{\mu\nu} 
\label{eq:energymomentumtensor}
 \end{eqnarray}
It reveals the collective gluons flow in the system. 
Therefore, we can temporarily summarize that the system 
represents an isotropic homogeneous  
perfect fluid of gluon dominated QGP. Further, to discuss the 
dynamics of glue lumps, it is also plausible 
to take the gluon as a field that independent of time, 
$\phi = \phi (x)$. If we compare Eq.(\ref{eq:energymomentumtensor}) 
with energy momentum tensor of ideal fluid 
$\cal{T}_{\mu\nu} = ({\cal E}$ $ + {\cal P}) u_\mu u_\nu$ - $ {\cal P}g_{\mu\nu}$, 
it becomes obvious that
\begin{eqnarray}
 {\cal E} = {\cal P} = 4 T g_F m_Q \phi.
 \label{eq:eos}
\end{eqnarray}
Here ${\cal P}$ and ${\cal E}$ denote the isotropic 
pressure and energy density for fluid field, respectively. 
Or, in currently discussed topic, it is the energy density 
and pressure distribution in the lump of gluon dominated QGP.

\section{Expression for distribution field}
\label{sec:field_equation}
The expression for the scalar field $\phi$ can be obtained by deriving the solution of Eq.~(\ref{eq:eq-of-motion-nonrel}). After applying the assumption
of the gluon homogeneity, the equation can be expressed as follow.  
\begin{equation}
\delta \partial_{tr} \phi + \gamma \partial_{rr} \phi + \xi + \beta \phi 
\partial_t \phi + \beta \phi \partial_r \phi - \lambda \partial_t \phi - \lambda  
\partial_r \phi = 0,
\label{eq:parsial-dif}
\end{equation}
with $\delta = \gamma |\mathbf{v}|$, $\xi = g_F \bar{Q} \gamma_\mu T Q$ , 
$\beta = i \gamma^2 |\mathbf{v}|^2$,  $\lambda = (g_G / g_F) A \gamma$, and $\gamma = (1-v^2)^{-\frac{1}{2}}$.
For a simplification but stay relevant, the pressure and energy density distribution of the lump of QGP is assumed to depend only to $r$, it is
 $\phi = \phi(r)$, and $\partial_r \rightarrow d_r$. So the term that involve the derivative of $t$ is vanished.
\begin{eqnarray}
\gamma d_{rr} \phi + \beta \phi d_r \phi - \lambda d_r \phi + \xi = 0 ,
\label{eq:nonhomogeneous_eq}
\end{eqnarray}
The solution for this equation is
\begin{eqnarray}
\phi &=& \Big( 2 \gamma \Big(  \frac{\exp[\frac{r \lambda}{2 \gamma}] \Lambda_1 [x]}{ 2 \gamma}  -    \frac{\exp[\frac{r \lambda}{2 \gamma}] \beta \xi \Lambda_2 [x]}{ 2^{1/3} \gamma^2 (- \frac{\beta \xi}{\gamma^2})^{2/3}} \nonumber\\
&&+  \Big( \frac{\exp[\frac{r \lambda}{2 \gamma}] \Lambda_3 [x]}{ 2 \gamma}  -    \frac{\exp[\frac{r \lambda}{2 \gamma}] \beta \xi \Lambda_4 [x]}{ 2^{1/3} \gamma^2 (- \frac{\beta \xi}{\gamma^2})^{2/3}}  \Big)  C_2  \Big)  \Big) \nonumber\\ 
&&\times 1 /  \Big( \beta \Big( \exp[\frac{r \lambda}{2 \gamma}] \Lambda_1 [x] +  \exp[\frac{r \lambda}{2 \gamma}] \Lambda_3 [x]  C_2 \Big)  \Big) .\nonumber\\
\end{eqnarray}
When we substitute $\delta,  \xi, \beta,  \lambda$, and $\gamma$ back to the equation, then $\phi$ appears as 
\begin{eqnarray}
\phi &=& \Big( \frac{2}{\sqrt{1-v^2}} \Big(  \frac{\exp[\frac{r g A}{2}] \Lambda_1 [x]}{ 2 / \sqrt{1-v^2}}  -    \frac{\exp[\frac{r g A}{2}]  (i v^2 \xi)^{1/3} \Lambda_2 [x]}{ 2^{1/3}} \nonumber\\
&&+  \Big( \frac{\exp[\frac{r g A}{2}] \Lambda_3 [x]}{ 2 / \sqrt{1-v^2}}  -    \frac{\exp[\frac{r g A}{2}]  (i v^2 \xi)^{1/3} \Lambda_4 [x]}{ 2^{1/3}}   \Big)  C_2  \Big)  \Big) \nonumber\\ 
&&\times 1 /  \Big( \frac{i v^2}{1 - v^2} \Big( \exp[\frac{r g A}{2}] \Lambda_1 [x] +  \exp[\frac{r g A}{2}] \Lambda_3 [x]  C_2 \Big)  \Big) .
\label{eq:phi}
\end{eqnarray}

Here,
\begin{eqnarray}
\Lambda_1 [x] &=& \frac{1}{3^{1/6} \Gamma \big[ \frac{2}{3} \big]} + \frac{3^{1/6} x}{\Gamma \big[ \frac{1}{3} \big]} + \frac{x^3}{6 \big( 3^{1/6} \Gamma \big[ \frac{2}{3} \big] \big)} + \frac{x^4}{4 \big( 3^{5/6} \Gamma \big[ \frac{1}{3} \big] \big)} + \mathcal{O} [x]^5 ~,\nonumber\\
\Lambda_2 [x] &=& \frac{3^{1/6}}{ \Gamma \big[ \frac{1}{3} \big]} + \frac{x^2}{2 \big( 3^{1/6} \Gamma \big[ \frac{2}{3} \big] \big)} + \frac{x^3}{ 3^{5/6} \Gamma \big[ \frac{1}{3} \big]} + \frac{x^5}{30 \big( 3^{1/6} \Gamma \big[ \frac{2}{3} \big] \big)} + \mathcal{O} [x]^6 ~,\nonumber\\
\Lambda_3 [x] &=& \frac{1}{3^{2/3} \Gamma \big[ \frac{2}{3} \big]} - \frac{x}{3^{1/3} \Gamma \big[ \frac{1}{3} \big]} + \frac{x^3}{6 \big( 3^{2/3} \Gamma \big[ \frac{2}{3} \big] \big)} - \frac{x^4}{12 \big( 3^{2/3} \Gamma \big[ \frac{2}{3} \big] \big)} + \mathcal{O} [x]^5 ~,\nonumber\\
\Lambda_4 [x] &=& -\frac{1}{3^{1/3} \Gamma \big[ \frac{1}{3} \big]} + \frac{x^2}{2 \big( 3^{2/3} \Gamma \big[ \frac{2}{3} \big] \big)} - \frac{x^3}{3 \big( 3^{1/3} \Gamma \big[ \frac{1}{3} \big] \big)} + \frac{x^5}{30 \big( 3^{2/3} \Gamma \big[ \frac{2}{3} \big] \big)} + \mathcal{O} [x]^6 ~, \nonumber
\end{eqnarray}
and
\begin{eqnarray}
x = \frac{-i 2 v^2 (\xi r + \frac{C_1}{ \sqrt{1-v^2}} ) + g^2 A^2}{2^{4/3} (-i v^2 \xi)^{2/3}} \nonumber
\end{eqnarray}
So far this non-trivial solution is well expressed. But to have a firm solution, some adjustment still need to be carry out at the future works. While for the rest, we will explore the expression for pressure and energy density in the system of gluon dominated quark-gluon plasma.

\begin{figure}[t]
 \centering
 \includegraphics[width=10cm]{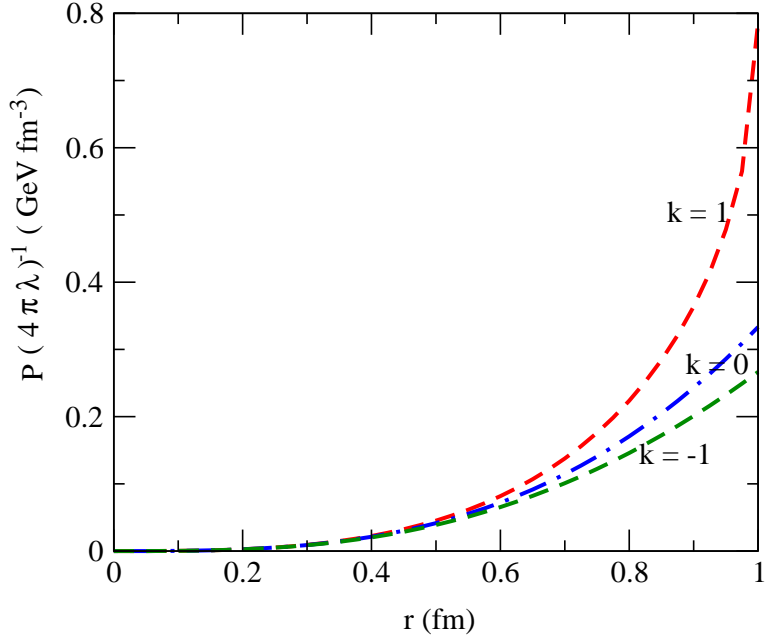}
 \caption{Pressure inside QGP-lump in the function of radius}
 \label{fig:qgpdensity}
\end{figure}

\section{Summary and Discussion}
\label{sec:summary_and_discussion}

Return to Eq.~(\ref{eq:eos}), ${\cal E}= {\cal P} = 4 T g_F m_Q \phi$,
the energy and pressure can be expressed as 
follows \cite{nugroho} 
\begin{eqnarray}
\rho  = P =  \int \mathcal{E} d^4 x =  \int {\cal P} d^4 x   = 4 T g_F f_Q m_Q \int \int  \phi dt dV ~,
\label{eq:density}
\end{eqnarray}
where $\mathcal{E}$ and $\mathcal{P}$ denote the isotropic 
pressure and density for single fluid field, respectively.\\
By adopting FRW geometry as the background and do integration on the spatial dimension, 
Eqs. (\ref{eq:density}) read 
\begin{eqnarray}
\rho  = P =  4 \pi \lambda \int R^3 \phi  dt \int \frac{r^2 dr}{\sqrt{1-kr^2}} .
\label{eq:rhof}
\end{eqnarray}
Here we have used 
$dV = {(R^3 r^2 sin\theta)}/{\sqrt{1- kr^2}} dr d\theta d\vartheta$, and 
$\lambda =  4 T g_F f_Q m_Q $.
Further, if we assign $\int^{2 \pi}_{0}  \int^{\pi}_{0}  \int dV = 4 \pi \int \frac{r^2 dr}{\sqrt{1-kr^2}} $ as $\zeta$, then the expression for $P$ and $\rho$ can be written as
\begin{equation}
P = \rho = \lambda \zeta \int R^3 \phi dt.
\end{equation}
In a case when $r$ is constant, $p = \rho \propto \phi (t) dt$, and also if  $\phi (t) \propto 1/t $, Then one arrive at  $P = \rho \propto \lambda \zeta R^3/t$. It indicates that the pressure and energy in the system decrease asymptotically following the increases of time.\\
Figure 1 shows the total pressure inside the QGP-lump as the function of radius. It is drawn for the simplest condition, where $\phi$ is assumed as a constant, and the scale factor $R = 1$. While $k = 1, 0, -1$ represent the space-time curvature that describe the close, flat, or open space-time that analogous to the close, flat, or open universe.  
One of the obvious property that appears here is the equation of state $P/\rho = 1$.
Such a ratio can indicates that in this model the QGP exist within radiation state. This estimate comes from the fact that in general the prerequisite condition for the radiation state is $P \sim 1/3 \rho$, and $P/\rho$ is getting smaller along the transition from radiation state to matter state.

Finally, the study revealed observables which should be accessible through heavy ion collisions experiments at the RHIC and LHC. With currently proposed calculation results, such experiments are in future expected to become a reference for adjustments and verifications of the dynamics theory of macroscopic 
behavior of QGP. 

\section*{Acknowledgments}

TPD thanks the Group for Theoretical and Computational Physics, 
Research Center for Physics, Indonesian Institute of Sciences (LIPI) 
for the warm hospitality during the completion of this work. 
T.P.D. and L.T.H. are 
supported by Riset Kompetitif LIPI under Contract 
No. 11.04/SK/KPPI/II/2016. \\


\section{References}
\begin{thebibliography}{0}
\bibitem{gottlieb} 
  Gottlieb S 2007
  {\it J.\ Phys.\ Conf.\ Ser.}\  {\bf 78}, 012023 
\bibitem{petreczky} 
  Petreczky P 2008 
  {\it Eur.\ Phys.\ J.} \ ST {\bf 155}, 123   
\bibitem{bouras} 
  Bouras I, Molnar E, Niemi H, Xu Z, El A, Fochler O, Greiner C, Rischke D H  2009 
  {\it Phys.\ Rev.\ Lett.}\  {\bf 103}, 032301   
\bibitem{romatschke} 
  Romatschke P 2010
  {\it Int.\ J.\ Mod.\ Phys.\ E} {\bf 19}, 1 
\bibitem{teaney} 
  Teaney D, Lauret J, Shuryak E V 2001
  {\it Phys.\ Rev.\ Lett.}\  {\bf 86}, 4783 
\bibitem{huovinen} 
  Huovinen P, Kolb P F, Heinz U W, Ruuskanen P V, Voloshin S A 2001
  {\it Phys.\ Lett.\ B} {\bf 503}, 58 
\bibitem{kolb} 
  Kolb P F, Heinz U W, Huovinen P, Eskola K J, Tuominen K 2001
  {\it Nucl.\ Phys.\ A} {\bf 696}, 197 
\bibitem{kolb1} 
  Kolb P F, Rapp R  2003
  {\it Phys.\ Rev.\ C} {\bf 67}, 044903 
\bibitem{hirano} 
  Hirano T, Tsuda K 2002
  {\it Phys.\ Rev.\ C} {\bf 66}, 054905 
\bibitem{baier} 
  Baier R, Romatschke P 2007
  {\it Eur.\ Phys.\ J.\ C} {\bf 51}, 677 (2007).
\bibitem{adams} 
  Adams J {\it et al.} 2003 
  {\it Phys.\ Rev.\ Lett.}\  {\bf 91}, 172302 
\bibitem{adare} 
  Adare A  {\it et al.} 2008  
  {\it Phys.\ Rev.\ Lett.}\  {\bf 101}, 232301 


\bibitem{tpdjun}
Djun T P, Handoko L T, Soegijono B, Mart T 2015
{\it Int. J. Mod. Phys. A }{\bf 30} 1550077 

\bibitem{sulaiman} 
 Sulaiman A, Fajarudin A,  Djun T P, Handoko L T 2009
{\it Int. J. Mod. Phys. A} {\bf 24}, 3630 

\bibitem{djun} 
  Djun T P, Handoko L T  2010  {\it in 
  Proceeding of the Conference in Honor of Murray Gell-Mann's 
  80th Birthday: Quantum Mechanics, Elementary Particles, Quantum Cosmology 
  \& Complexity, edited by H. Fritzsch and K. K. Phua 
  (Nanyang Technological University, Singapore) }, p. 419 
  arXiv:1109.6066 [hep-ph].

\bibitem{nugroho} 
  Nugroho C S,  Latief A O, Djun T P, Handoko L T 2012
  {\it Grav. Cosmol.} {\bf 18}, 32 

  
  
\bibitem{hobson}
  Hobson M P, Efstathiou G, and  Lasenby A N (Cambridge University Press 2006)
  {\it General Relativity} 
  

\end{thebibliography}
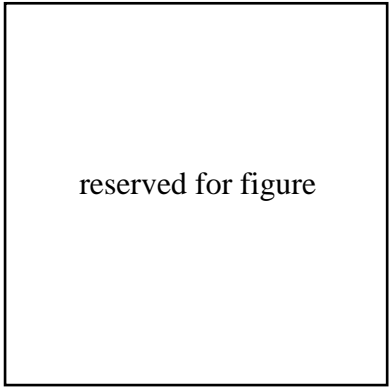
\end{document}